\documentclass[final,5p,times,twocolumn]{elsarticle}
\usepackage{multicol,caption}
\usepackage{subcaption}
\usepackage{graphicx}
\usepackage{bbm}
\usepackage{xspace}
\usepackage{amsfonts}
\usepackage{siunitx}
\usepackage{color}
\usepackage{graphics}
\usepackage{amsmath}
\usepackage{blindtext}
\journal{Elsevier}

\begin{document}

\begin{frontmatter}

\title{Multi-fidelity Gaussian process based empirical potential development for Si:H nanowires}

\author[label1]{Moonseop Kim}
\address[label1]{School of Mechanical Engineering, Purdue University, West Lafayette, IN 47906-2045}

\author[label2]{Huayi Yin}
\address[label2]{School of Computer and Information Engineering, Xiamen University of Technology, Xiamen, Fujian 361024, China}

\cortext[cor1]{Corresponding author}
\address[label3]{Department of Mathematics, Purdue University, West Lafayette, IN 47906-2045}
\author[label1,label3]{Guang Lin\corref{cor1}}
\ead{guanglin@purdue.edu}

\begin{abstract}
In material modeling, the calculation speed using the empirical potentials is fast compared to the first principle calculations, but the results are not as accurate as of the first principle calculations. First principle calculations are accurate but slow and very expensive to calculate. In this work, first, the H-H binding energy and H$_2$-H$_2$ interaction energy are calculated using the first principle calculations which can be applied to the Tersoff empirical potential. Second, the H-H parameters are estimated. After fitting H-H parameters, the mechanical properties are obtained. Finally, to integrate both the low-fidelity empirical potential data and the data from the high-fidelity first-principle calculations, the multi-fidelity Gaussian process regression is employed to predict the H-H binding energy and the H$_2$-H$_2$ interaction energy. Numerical results demonstrate the accuracy of the developed empirical potentials.

\end{abstract}


\end{frontmatter}


In the last three decades, empirical potentials have been advanced. With the advance of supercomputers, these potentials are anticipated to be widely used for the next three decades \cite{1}. Atomistic calculations by empirical potentials can be utilized in understanding the structural aspects of Si or Si-H systems that are found in many important areas such as the surface of nano-patterning Si \cite{2,3}, nano-electro-mechanical systems (NEMS) \cite{4}, superconductivity of silane \cite{5}, optical modulators \cite{6}, and applications of $\alpha$-Si:H materials \cite{7}. In the past, empirical potentials for Si \cite{8,9,10,11} and for Si-H \cite{12,13,14} have been developed. But the bulk elastic properties of Si cannot be resolved using such empirical potentials. 
In [12], it has been shown that at the hydrogen-induced reconstruction of the silicon surface, the distance between hydrogen and hydrogen is 1.64\AA \, and bond angle H-Si-H is 106$^{\circ}$ using existing empirical potential. However, when H-H distance and the bond angle are compared with the results from the first-principle calculations, H-H distance and the bond angle are 2.1638\AA \, and 104.805$^{\circ}$ respectively.
In this situation, the bond angle is distinguished from 1.195$^{\circ}$, which means that the difference of the bond angles can be ignored, however, the biggest issue is that H-H distance is distinguished from 0.5238\AA. Hence, if the existing empirical potential is used for Si nanowires, the computation speed is fast, but the results obtained from the existing empirical potentials are not accurate compared to the results from the first-principle calculations. It is critical to fix such errors. In this paper, we propose two novel techniques to construct the empirical potentials with an emphasis on parameter fitting and multi-fidelity modeling, in which the relationship between material properties and potential parameters is explained. The input database has been obtained from the density functional theory (DFT) calculations with the Vienna ab initio simulation package (VASP) \cite{15}. This paper is constituted as follows. First, the structure of silicon nanowires passivated hydrogen is introduced with some specific shapes. Second, governing equations of Tersoff empirical potentials are presented to explain which parameters can be obtained from H-H binding energy and H$_2$-H$_2$ interaction energy. Third, we give a brief explanation of the multi-fidelity Gaussian process regression for prediction of the results of H-H binding energy and H$_2$-H$_2$ interaction energy. Fourth, we represent three optimization methods for H-H parameter fitting. The root-mean-square-error obtained by the Nelder-Mead Simplex Method is compared with the results from the other two optimization methods. Lastly, we evaluate the mechanical properties (Young's modulus and equilibrium elongation) using the estimated parameters obtained from the H-H parameter fitting.

\begin{figure}
\centering
\includegraphics[scale=0.5]{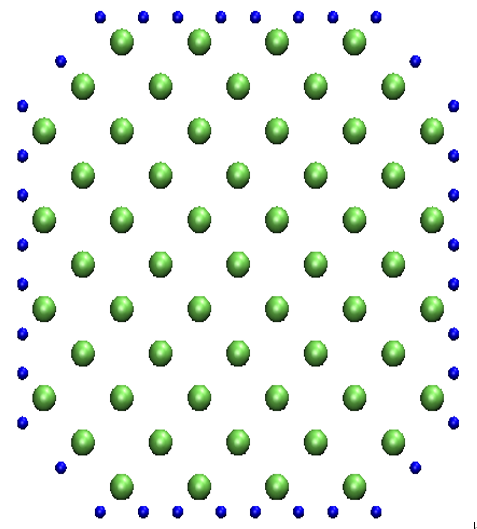}
\caption{Cross-section of silicon nanowires passivated hydrogen $<$001$>$.}
\label{fig:1}       
\end{figure}

\indent In this study, Si nanowires passivated hydrogen model is chosen. If Si nanowires have dangling bonds, it will oxidize in the air circumstance. By passivating hydrogen to the surface of Si nanowires, it can be stabilized from the oxidization. Fig. 1 \cite{16} has expressed a cross-section of silicon nanowires passivated hydrogen. Green dots and blue dots represent silicon atoms and hydrogen atoms, respectively. Cross-section of Si nanowires represents by the Wulff structure selected by minimizing the surface energy. It can be stabilized by passivating hydrogen to the surface of Si nanowires. In the mechanical property, Young's modulus is calculated after the H-H parameter fitting by increasing the size of cross-section compared to the results of Young's modulus using the existing empirical potentials and the first-principles calculations.
\begin{figure}[t]
\includegraphics[scale=0.55]{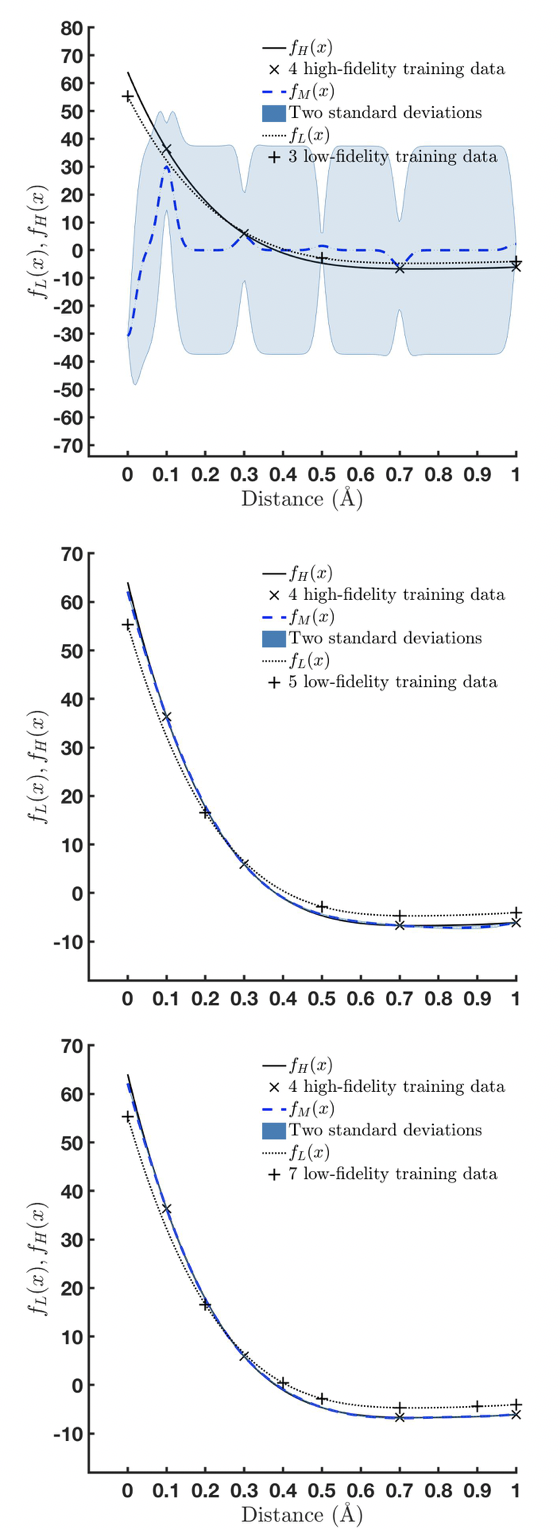}
\centering
\caption{Multi-fidelity prediction results of H-H binding energy with high-fidelity samples ($N_H$=4) and three different numbers of low-fidelity samples ($N_L$=3, 5, 7). If we increase $N_L$, the accuracy is increased (the standard deviation of the high-ﬁdelity decreases). }
\end{figure}

\indent The atomistic computer simulations based on the empirical potential is fast for calculation. In this system, the number of atoms is not limited compared to the first principles, however, the accuracy of calculation is not adequate, therefore, reliability of the empirical potential presented so far is needed to verify. Various empirical potentials depend on the material, for instance, Nickel (Ni) and Titanium (Ti) are calculated through EAM (Embedded Atom Method) \cite{17}. and Silicon (Si) is calculated through Tersoff empirical potential \cite{9} and Stillinger-Weber empirical potential \cite{8}. In this study, silicon nanowires passivated hydrogen model, and Tersoff empirical potential are used to verify the accuracy of existing Tersoff empirical potentials. Tersoff empirical potential is based on the concept of bond order, the force of bonds between atoms is not consistent and depends on the local environment. The total energy function is given as \cite{12}

\begin{equation} \label{eqn1}
V=0.5*\sum_{i,j, i\neq j}f_R(r)+b_{ij}f_A(r)
\end{equation}

\begin{equation} \label{eqn2}
f_R(r)=Aexp(-\lambda_1r_{ij})
\end{equation}

\begin{equation} \label{eqn3}
f_A(r)=-Bexp(-\lambda_2r_{ij})
\end{equation}

\begin{equation} \label{eqn4}
b_{ij}=(1+\zeta_{ij}^\eta)^{-\delta}
\end{equation}
V is total energy function, $f_R(r)$ and $f_A(r)$ are repulsive energy and attractive energy respectively. These functions are defined as function of distance between i and j atoms, r is interatomic distance and $b_{ij}$ is bond order. In this study, A, B, $\lambda_1$ and $\lambda_2$ are decided as H - H fitting parameters.
\begin{multline} \label{eqn5}
\zeta_{ij}=\sum_{k\neq i,j}f_c(r_{ik})[c+d\{H(N)-cos\theta_{ijk}\}^2] \\
\times exp[\alpha\{(r_{ij}-R_{ij}^{(e)})-(r_{ik}-R_{ik}^{(e)})^\beta]
\end{multline}

\noindent$\zeta_{ij}$ is the function of effective coordination number, H(N) is the function of bond number, cos$\theta_{ijk}$ is the bond angle, $r_{ij}$ and $r_{ik}$ are the distance between i and j atoms and between i and k respectively and $R_{ij}^{(e)}$ and $R_{ik}^{(e)}$ are equilibrium distance between i and j atoms and between i and k respectively. Lastly, in this study, $\alpha$, $\beta$, $\eta$, $\delta$ and c are determined from H$_2$-H$_2$ parameter fitting.

\begin{equation} \label{eqn6}
  f_c(n) = \left \{
  \begin{aligned}
    &1 \\
    &\frac{1}{2}-\frac{9}{16}sin(\pi\frac{r_{ij}-R}{2D})-\frac{9}{16}sin(3\pi\frac{r_{ij}-R}{2D}) \\
    &0
  \end{aligned} \right.
\end{equation}

\begin{figure}[t]
\includegraphics[scale=0.54]{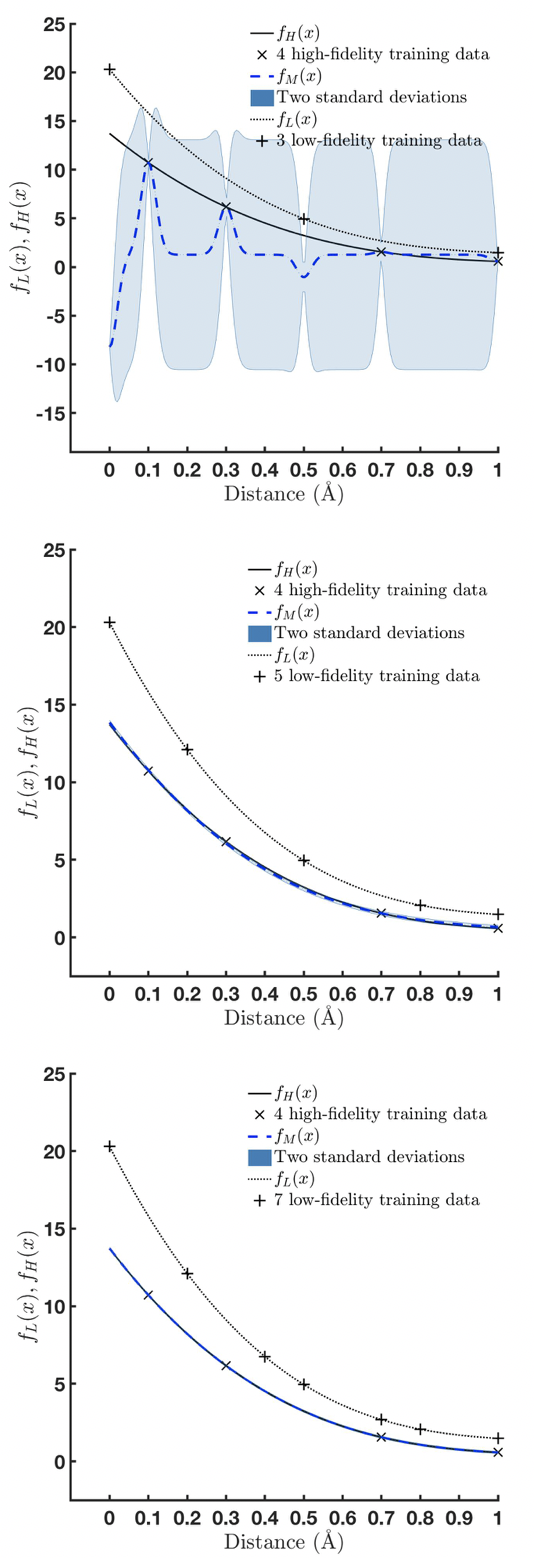}
\centering
\caption{Multi-fidelity prediction results of H$_2$-H$_2$ interaction energy with high-fidelity samples ($N_H$=4) and three different numbers of low-fidelity samples ($N_L$=3, 5, 7). If we increase $N_L$, the accuracy is increased (the standard deviation of the high-ﬁdelity decreases). }
\end{figure}

\noindent$f_c(n)$ is a cutoff function determining whether there is coherence or not between the atom and its neighbor atom. r is the interatomic distance, R and D are constants to determine the range of the inter-atomic influence. if the interatomic distance is less than R - D, the influence is 1. if the interatomic distance is larger than R - D, the influence is 0. Finally, if the interatomic distance is between R-D and R+D, it is influenced by Eq.(5).

\indent Here, we provide the steps for multi-fidelity modeling with Gaussian processes (GP). The steps on multi-fidelity are given as 

\begin{equation} \label{eqn7}
u_1(x)\thicksim\mathcal{GP}(0, k_1(x,x^{'};\theta_1))
\end{equation}

\begin{equation} \label{eqn8}
u_2(x)\thicksim\mathcal{GP}(0, k_2(x,x^{'};\theta_2))
\end{equation}

\noindent $u_1(x)$ and $u_2(x)$ are independent. In the Gaussian process regression, it is assumed that the mean of $\mathcal{GP}$ is zero and k(x,x';$\theta$) is the covariance matrix between all possible pairs (x, x') in the set of vectors of hyper-parameters $\theta$. As shown in \cite{18}, the basic idea is that we begin with two independent $\mathcal{GP}$ $u_1(x)$ and $u_2(x)$; then we define the low-fidelity and the high-fidelity models: \cite{19,20,21,22,23,24}

\begin{equation} \label{eqn9}
f_L(x)=u_1(x)
\end{equation}

\begin{equation} \label{eqn10}
f_H(x)=\rho u_1(x)+u_2(x)
\end{equation}

\noindent This demonstrates the ``relationship" between the low- and high-fidelity models since both include the $\mathcal{GP}$ $u_1(x)$.  In particular, setting $k_1$=cov[$u_1$, $u_1$] and $k_2$=cov[$u_2$, $u_2$] we have:

\begin{equation} \label{eqn11}
K_{LL}=cov[f_L,f_L]=cov[u_1,u_1]=k_1
\end{equation}

\begin{multline} \label{eqn12}
K_{LH}=cov[f_L,f_H]=cov[\rho u_1+u_2,u_1] \\
=\rho cov[u_1,u_1]+cov[u_2,u_1]=\rho k_1
\end{multline}

\begin{multline} \label{eqn13}
K_{HH}=cov[f_H,f_H]=cov[\rho u_1+u_2,\rho u_1+u_2] \\
=\rho^2 cov[u_1,u_1]+\rho cov[u_2,u_1] \\
+\rho cov[u_1,u_2]+\rho cov[u_2,u_2]=\rho^2k_1+k_2
\end{multline}

\noindent cov[$u_1$, $u_2$] = 0 and cov[$u_2$, $u_1$] = 0 by independence and to sum up $K_{LL}$,$K_{LH}$,$K_{HH}$:

\begin{equation} \label{eqn14}
\begin{split}
K_{LL}=k_1(x,x^{'};\theta_1)\\
K_{LH}=\rho k_1(x,x^{'};\theta_1)\\
K_{HH}=\rho^{2} k_1(x,x^{'};\theta_1)+k_2(x,x^{'};\theta_2)\\
\end{split}
\end{equation}

\noindent This gives us a complete model that incorporates both the low- and high-fidelity. In particular, we model the column vector [$f_L(x)$; $f_H(x)$] using a zero-mean prior and the covariance matrix defined block-wise by: [ $K_{LL}$, $K_{LH}$; $K_{HL}$, $K_{HH}$]. Since the mean and covariance are known, the whole Gaussian process model is specified, and the training can be performed using the standard procedure.

\begin{equation} \label{eqn15}
\begin{bmatrix} f_L(x) \\ f_H(x) \end{bmatrix}
 \thicksim\mathcal{GP}\Bigg(0,
  \begin{bmatrix}
   K_{LL}(x,x^{'}) &
   K_{LH}(x,x^{'}) \\
   K_{HL}(x,x^{'}) &
   K_{HH}(x,x^{'})  
   \end{bmatrix}\Bigg)
\end{equation}

\noindent In training, based on $\{$$x_L$, $y_L$$\}$, $\{$$x_H$, $y_H$$\}$, $N_L$ $<$$<$ $N_H$,

\begin{multline} \label{eqn16}
\begin{bmatrix} y_L(x) \\ y_H(x) \end{bmatrix}
 \thicksim \\ \mathcal{N} \Bigg(0,
  \begin{bmatrix}
   K_{LL}(x_L,x_L)+\sigma_L^2I &
   K_{LH}(x_L,x_H) \\
   K_{HL}(x_H,x_L) &
   K_{LL}(x_H,x_H)+\sigma_H^2I  
   \end{bmatrix}\Bigg)
\end{multline}

\noindent To represent the uncertainty (or noise) in the observation data, the covariance of the noise for both the low- and high-fidelity data is added on the diagonal of the covariance matrix in Eq. 16. The level of the noise in the observation data will affect the prediction accuracy as shown in Fig. 2 and 3.

\begin{multline} \label{eqn17}
NLML(\theta_1,\theta_2,\rho)= \\ \frac{1}{2}y^T\mathcal{K}^{-1}y+\frac{1}{2}log|\mathcal{K}|
+\frac{N_L+N_H}{2}log(2\pi)
\end{multline}

\noindent $\mathcal{K}$ is the covariance matrix and the Negative Log Marginal Likelihood (NLML) is used as the "cost function" which should be minimized to get the best-fit model by using hyper-parameters $\theta$. In prediction, if we consider a Gaussian likelihood and the posterior distribution is easy to apply and can be used to involve predictive deduction for a new output $f_H$, given a new input $x^*$ as

\begin{multline} \label{eqn18}
f_H(x^{*}|y\thicksim\mathcal{N}([K_{HL}(x^{*},x_L)K_{HH}(x^{*},x_H)]\mathcal{K}^{-1}y,K_{HH}(x^*,x^*) \\
-K(x^*,x)\mathcal{K}^{-1}K(x,x^*))
\end{multline}

\noindent In numerical results, to reduce the computational cost for expensive calculations (DFT), Multi-fidelity Gaussian process regression for prediction \cite{19,20,21,22,23,24} is used. H-H binding energy and H$_2$-H$_2$ interaction energy obtained from the empirical potential are applied to the low-fidelity model. Results of H-H binding energy and H$_2$-H$_2$ interaction energy obtained from DFT are implied to the high-fidelity model with a limited number of samples due to high computational cost.

\noindent In Fig. 2 and 3, the number of high-fidelity samples ($N_H$=4) is fixed and we compare the standard deviation of the high-fidelity prediction by increasing the number of low-fidelity samples. As we can see, the standard deviation of the high-fidelity is decreased when the number of low-fidelity samples is increased.

\begin{table*}[ht]
\caption{To fit H - H parameters for molecular hydrogen, three optimization methods are compared which are Nelder-Mead Simplex Method (N - M) \cite{25}, Broyden-Fletcher-Goldfarb-Shanno Quasi-Newton Method (BFGS) \cite{26,27,28,29}, Trust-Region Method (T - R) \cite{30,31}.}
\centering
\begin{tabular}{ c| c c c c } 
 \hline
 \hline
   & N-M(This work) & BFGS & T-R & Existing \\
 \hline 
 A & 87.5482 & 87.5471 & 80.0752 & 80.07 \\ 
 B & 18.2497 & 18.2498 & 31.3714 & 31.38 \\ 
 $\lambda_1$ & 5.2898 & 5.2898 & 3.9932 & 4.2075 \\ 
 $\lambda_2$ & 1.0290 & 1.0290 & 1.4134 &	1.7956 \\ 
 $R^{(e)}$ &  0.75 & 0.75 & 0.75 & 0.74 \\ 
 R & 3.5 & 3.5 & 3.5 & 1.4 \\ 
 D & 0.5 & 0.5 & 0.5 & 0.3 \\ 
 h & 0 & 0 & 0 & 0 \\ 
 d & 0 & 0 & 0 & 0\\ 
 $\alpha$ & 3 & 3 & 3 & 3 \\ 
 $\beta$ & 1 & 1 & 1 & 1 \\ 
 c & 1.3328 & 1.3322 & 3.9550 &	4 \\ 
 $\delta$ & 0.6923 & 0.6926 &	0.3576 & 0.80469 \\ 
 $\eta$ & 1.2866 & 1.2865 & 0.8464 & 1 \\ 
 RSME (Binding energy) & 0.0014 & 0.0014 & 0.1341 \\ 
 RSME (Interaction energy) & 0.0012 & 0.0013 & 0.0521 \\ 
 \hline
\end{tabular}
\end{table*}

The structure of Si nanowires is passivated by hydrogen to prevent oxidation. We divide the Si nanowires into three parts for parameter fitting. First, Si-Si parameter fitting is needed for the internal structure of Si nanowires, Second, H-H parameter fitting is required for the surface computation. Lastly, Si-H parameters for the interface calculation are needed. In this study, the H-H parameters fitting is focused on using H-H binding energy and H$_2$-H$_2$ interaction energy for surface computation. In the surface of Si nanowires, there are two forces for hydrogen relations that are attractive and repulsive forces. This H-H parameter fitting is complicated so we have to design systematically. When we fit H-H parameters using H-H binding energy, we only use Eq. (1-3) except for Eq. (4 and 5). This is because H-H binding energy is calculated by two hydrogen atoms which can be neglected $\zeta_{ij}$ term that needs more than three atoms, however, H$_2$-H$_2$ parameter fitting by using interaction energy is applied to more than three hydrogen atoms so it should be applied to $\zeta_{ij}$ term. In Table. 1, three methods of optimization were used for H-H parameters fitting. Finally, we investigate the error of each method between DFT results and the fitting line through the root-mean-square-error in Table 1. The Nelder-Mead simplex method is the best compared to the other two optimization methods. The parameters using the Nelder-Mead simplex method were chosen, which are listed in Table.1.

\begin{figure}[t]
\includegraphics[scale=0.5]{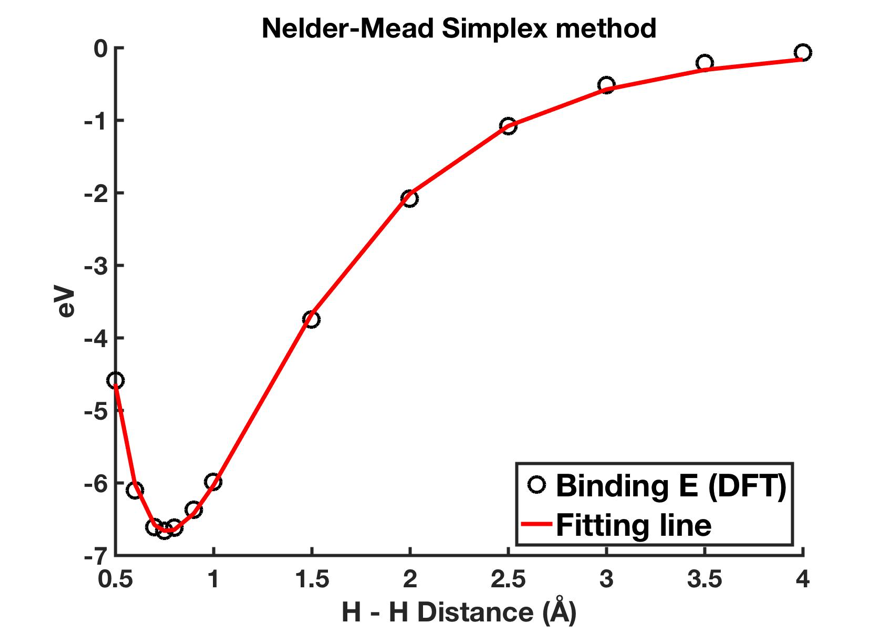}
\centering
\caption{H - H parameter fitting using the Nelder-Mead Simplex method.}
\end{figure}

In the study, we compute the binding energy using DFT for reference results that are adjusted to the results of the empirical potentials objective function. Parameters can be obtained after fitting between hydrogen and hydrogen. In Fig. 4, black dots represent DFT results for H-H binding energy and the red line is the fitting line. A, B, $\lambda_1$, $\lambda_2$ and R$^{(e)}$ values are listed on Table.1.

\begin{figure}[t]
\includegraphics[scale=0.5]{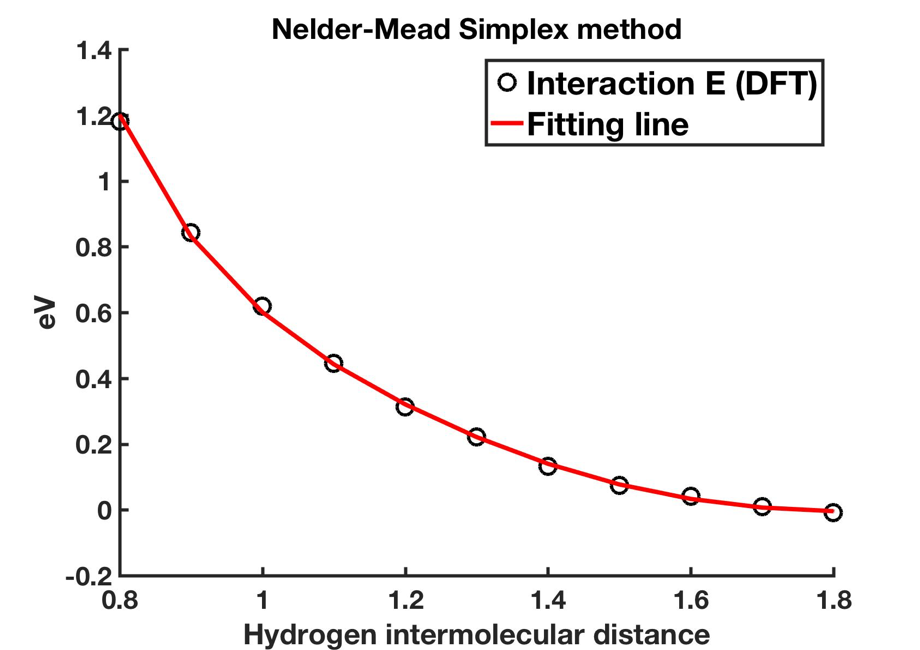}
\centering
\caption{H$_2$-H$_2$ parameter fitting using the Nelder-Mead Simplex method.}
\end{figure}

We compute the interaction energy using DFT for reference results that are adjusted to the results of the empirical potentials objective function and we obtained parameters after fitting hydrogen inter-molecular. We use the Nelder-Mead Simplex methods of optimization for the H - H parameter fitting using the interaction energy. In Fig. 5, the black dots represent the DFT results for H$_2$-H$_2$ interaction energy, and the red line is the fitting line. In this H - H fitting, A, B, $\lambda_1$, $\lambda_2$ and R$^{(e)}$ are obtained from H-H parameter fitting using the binding energy applied to H$_2$-H$_2$ parameter fitting and $\alpha$, $\beta$, $\eta$, $\delta$ and c values are presented in Table 1.

\begin{figure}[t]
\includegraphics[width=8cm,height=7cm]{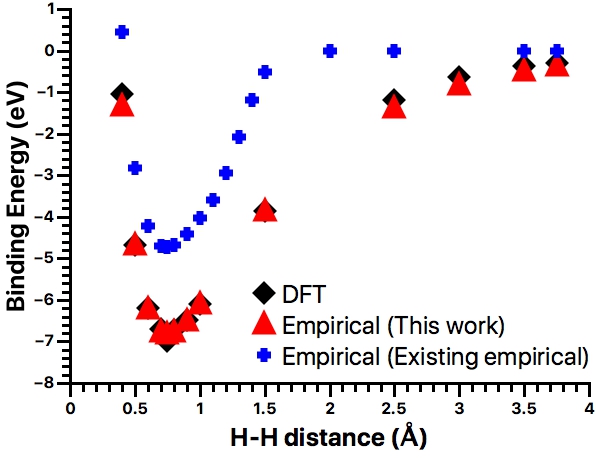}
\centering
\caption{Binding energy after H-H parameter fitting.}
\end{figure}

\begin{figure}[t]
\includegraphics[width=8cm,height=7cm]{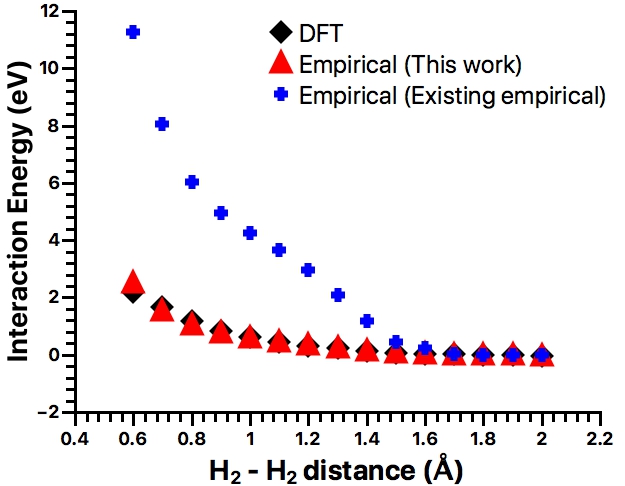}
\centering
\caption{Interaction energy after H-H parameter fitting.}
\end{figure}


Fig. 6 and 7 represent the H-H binding energy and H$_2$-H$_2$ interaction energy compared to DFT, empirical potential (this work), and existing empirical potential. In Fig. 6 and 7, cross-data represents the H-H binding energy and H$_2$-H$_2$ interaction energy using the existing Tersoff empirical potential parameters. It is critical to note that these data’s curve shape is not smooth because, in existing classical MD, the cutoff range is too short to calculation, so it does not calculate for the long-range interaction. In this study, we fixed cut off range much longer and calculate the long-range of hydrogen molecules. As we can see, DFT and empirical potential (this work) result matches each other after H-H parameters fitting. We have developed a systematic process to construct empirical potential for Si nanowires' passivated hydrogen.

\begin{figure}[t]
\includegraphics[width=8cm,height=7cm]{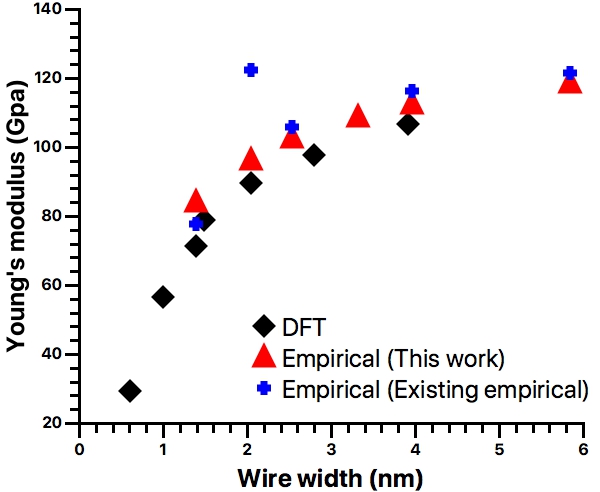}
\centering
\caption{Young's modulus increasing as wire width of Si nanowires increasing.}
\end{figure}

\begin{figure}[t]
\includegraphics[width=8cm,height=7cm]{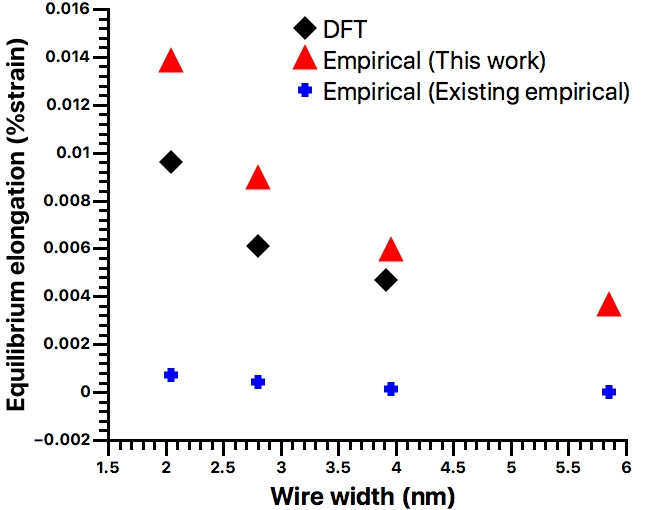}
\centering
\caption{Equilibrium elongation increasing as wire width of Si nanowires increasing.}
\end{figure}

In Figs. 8 and 9, mechanical properties were performed using H-H parameters which are obtained after H-H part parameter fitting. To evaluate the new fit of the H-H part, Young's modulus and equilibrium elongation of Si nanowires are calculated by increasing the wire width of Si nanowires. We compared the numerical results with the DFT results and existing empirical results in Figs. 8 and 9. The size reliance of Young's modulus and equilibrium elongation shows critical improvement compared to the DFT results and the existing potential results. Until now, the surface part of Si nanowires is fitted using our systematic fitting method and shows mechanical properties to prove enhancement. However, the improvement of the irregular mechanical properties can be observed by the H-H parameter fitting of the surface. But the perfect result of matching could not be observed. The reason for this is that not only the surface but also the silicon-hydrogen parameter fitting between the silicon and the surface must be performed. In addition, the hydrogen parameters obtained so far are limited to the calculation of the mechanical properties of nanowires with hydrogen and silicon. Furthermore, it is necessary to derive parameter fitting that can be applied to various types of materials, and potential errors of existing empirical potentials must be corrected for calculations using various materials as well as silicon nanowires.

\section*{Acknowledgments}
We gratefully acknowledge the support from the National Science Foundation (DMS-1555072 and DMS-1736364).



\bibliographystyle{elsarticle-num}
\bibliography{sample}

\end{document}